\documentstyle[eqsecnum,aps,prd]{revtex}

\newtheorem{theorem}     {Theorem}

\newtheorem{definition}  {Definition}

\newtheorem{corollary}   {Corollary}
{{\hspace*{1em}\hfill{$\Box$}}\\}
{{\hspace*{1em} \hfill{$\Box$}}}

\begin{document}

\title{Asymptotic singular behavior of 
Gowdy spacetimes in string theory}

\author{Makoto Narita\thanks{narita@se.rikkyo.ac.jp}, 
Takashi Torii\thanks{torii@resceu.s.u-tokyo.ac.jp}\thanks{
Presently at Research Center for the Early Universe, 
University of Tokyo, Hongo, Bunkyo, Tokyo, 113-0033, Japan and 
Advanced Research Institute for Science and Engineering, Department of Physics, 
Waseda University, Ohkubo, Shinjuku, Tokyo, 169-8555, Japan}, 
and Kengo Maeda\thanks{g\_maeda@amethyst.gravity.phys.waseda.ac.jp}}

\address{* Department of Physics, Rikkyo University, Nishi-Ikebukuro, Toshima, 
Tokyo 171-8501, Japan\\
\dag Department of Physics, Tokyo Institute of Technology, 
Ohokayama, Meguro, 
Tokyo 152-8551, Japan\\
\ddag Department of Physics, Waseda University, Ohkubo, 
Shinjuku, Tokyo 169-8555, 
Japan}
\maketitle

\begin{abstract}
We study $T^{3}$ 
Gowdy spacetimes in the Einstein-Maxwell-dilaton-axion system
and show by the Fuchsian algorithm that they have in general
asymptotically velocity-term dominated singularities.  
The families of the corresponding solutions depend on the maximum number of 
arbitrary functions.  
Although coupling of the dilaton field with the Maxwell and/or the axion 
fields corresponds to the ``potential'' which appears in the 
Hamiltonian of vacuum Bianchi IX spacetimes,
our result means that the spacetimes do not become 
Mixmaster necessarily.
\end{abstract}

\section{Introduction}
\label{intro}
The singularity theorem gives us some sufficient conditions for the existence 
of spacetime singularities (a spacetime with incomplete geodesics)
~\cite{HP,HE}.  
These conditions fall into four categories : (1) the strong energy condition; 
(2) the generic condition; (3) the chronology condition and (4) 
the existence of trapped regions.  
In addition, 
after the discovery of the singularity theorem, 
various generalization by replacing these conditions 
have been suggested by 
some authors (see e.g. Refs.\cite{S,MIN,N}).  
So, although, we have new many variations of the singularity theorem, 
none of them gives information on the nature of the 
singularity.  

To the best of our knowledge, the first research about the nature  of the 
singularity was done by Belinskii, Khalatnikov and Lifshitz (BKL)
~\cite{BKL}.  
They investigated how the spacetime evolves into the singularity
and conjectured that the dynamics of nearby observers would decouple near 
singularities.  (the {\it BKL conjecture}.)  
BKL described a vacuum and spatially homogeneous 
(mainly Bianchi IX) spacetime 
singularity as an infinite sequence of Kasner epochs (``oscillatory approach 
to the singularity'').  
Independently, 
Misner showed the same behavior in terms of exponentially growing 
``potential'' terms constructed by the spatial curvature.
The spacetime is bounced by the potentials a infinite number of times.
Such behavior is called the {\it Mixmaster dynamics}~\cite{CM}.  
Then, BKL further speculated that 
a generic singularity should be locally behaved like a Mixmaster type. 
To date, the BKL conjecture has neither been validated nor invalidated {\it in 
general} by rigorous arguments 
although Ringstr\"{o}m has recently have a result which supports the validity 
of the BKL conjecture, that is, he has 
shown rigorously that the Bianchi type IX solutions 
converge to an attractor consisting of the closure of the vacuum type II 
orbits\cite{HR}.  

There is a special case of the BKL conjecture called 
the {\it asymptotically velocity-term dominated} (AVTD) singularity. 
It is not described by a infinite sequence of Kasner epochs but
by only one epoch (i.e., ``no oscillatory approach 
to the singularity'')~\cite{IM,WE}.  
In brief, the characteristic feature of the AVTD singular behavior is that 
all spatial derivative terms of the field equations are negligible
sufficiently close to the singularity. 
Hence, Kasner spacetimes are necessarily (A)VTD.  
It is possible to make rigorous arguments on the nature of 
the AVTD singularity 
since the AVTD solutions are simpler than the Mixmaster solutions in the 
sense that there is no complicated oscillation.  
Our interest is to verify BKL conjecture and clarify 
whether spacetimes is AVTD or not in non-vacuum and/or spatially inhomogeneous 
cases.

In order to attack the above issue, we will consider
Gowdy spacetimes~\cite{G}.  They are spatially compact spacetimes which have 
$U(1)\times U(1)$ symmetry and vanishing twist.
Gowdy spacetimes are adopted in various investigations 
because they are 
one of the most manageable inhomogeneous spacetimes
and can give essential features of inhomogeneity in spite
of its simpleness. 
The behavior near the singularity in Gowdy spacetime
have also been discussed in various situations.
In the vacuum case, Isenberg and Moncrief have shown that 
polarized Gowdy spacetimes are AVTD~\cite{IM}.  
Grubi\v{s}i\'{c} and Moncrief~\cite{GM} and 
Kichenassamy and Rendall~\cite{KR}
have shown that  non-polarized 
$T^{3}$ Gowdy spacetimes have AVTD 
singularities in generic in the sense that the AVTD singular solutions to 
Einstein's equations depend on the maximal number of arbitrary functions 
if the {\it low velocity condition} 
is satisfied.
When one of these functions is constant (i.e. non-generic), it was 
also shown 
without the low velocity condition 
that non-polarized $T^{3}$ Gowdy spacetimes are AVTD~\cite{KR}.  
Using numerical method, 
Berger, Garfinkle and  Moncrief found that 
non-polarized 
$T^{3}$ Gowdy spacetimes have AVTD singularities even in
the situation that the low velocity condition is broken 
initially~\cite{BM,BG,BB98}.  
Recently, Garfinkle has shown that 
$S^{2}\times S^{1}$ Gowdy 
spacetimes have AVTD singularities numerically~\cite{DG}.  

AVTD solutions with matter fields present have been found.  
Such examples include 
polarized Gowdy spacetimes 
admitting scalar fields minimally coupled with 
the Maxwell field~\cite{CCM} and the Einstein-dilaton-axion (EDA) 
system in polarized Gowdy 
spacetimes~\cite{JBK}.  
It has been shown numerically, however,
that magnetic Gowdy spacetimes are not AVTD but Mixmaster~\cite{BB98,WIB}.  
Hence, it is nontrivial to determine whether the spacetime is Mixmaster
or AVTD type.  
Because of these facts, we shall study the non-vacuum and 
spatially inhomogeneous case.

There is an another point which needs to be clarified.  
Belinskii and Khalatnikov have suggested that the existence of massless 
scalar fields suppresses Mixmaster behavior by examining the Bianchi I 
spacetimes~\cite{BK}.  
It is due to the fact that the scalar field is algebraically equivalent
to the stiff matter in some cases.
Recently, Berger found some confirmation for 
this suggestion by numerical analysis~\cite{BB00}.  
In both papers, however, it was suggested that exponential coupling
of the scalar field to the Maxwell field 
restores Mixmaster behavior which was suppressed initially by 
the scalar field~\cite{BK,BB00}.  
It is important to
note that such matter fields arises naturally from low energy effective 
superstring theory~\cite{STW}, i.e. the Einstein-Maxwell-dilaton-axion 
(EMDA) system. 
The EMDA system has been discussed actively in the context of the
black hole solutions and singularities, and gives characteristic features 
different from the Einstein-Maxwell (EM) system~\cite{MTN}.

Our purpose in this paper is to explore the nature of 
singularities in the EMDA system with 
Gowdy spacetimes on $T^{3}\times {\bf R}$. 
We make use of the
Fuchsian algorithm developed by Kichenassamy and Rendall~\cite{KR}.  
Although the system treated here is very complicated, 
the Fuchsian algorithm is powerful enough for the system 
since this algorithm is independent of non-linearity, number of functions 
and dimension of space.  
We will show that there exist families of AVTD singular solutions of the EMDA 
field equations, which are generic in the sense that they depend 
on the maximal number of arbitrary functions.  
Hence, our result shows that exponential coupling of the
scalar field to the Maxwell field does not necessarily
lead the Mixmaster behavior.  

Organization of this paper is as follows.  
In Sec.~\ref{avtd}, an appropriate definition of AVTD is given.  
In Sec.~\ref{why-emda}, we discuss the importance of studying the EMDA 
system.  
Sec.~\ref{emda-gowdy} is devoted to introducing $T^{3}$ Gowdy 
spacetimes in the EMDA system.  
Sec.~\ref{Fuchsian algorithm} presents a brief review of the Fuchsian 
algorithm.  
Our main results are given in Sec.~\ref{results}.  
Finally, Sec.~\ref{summary} is a summary.  

\section{Asymptotically velocity-term dominated (AVTD) singularity}
\label{avtd}
In this section, we give precise definition of AVTD.
In contrast to the Mixmaster singularity which is complicated, 
the AVTD singularity is called simple or quiescent~\cite{ELS,IM}.  
An essential difference between Mixmaster and AVTD singularities is whether 
spatial curvature terms in Einstein's equations becomes dominant or not 
as the system approaches the 
singularities.  
As mentioned in the previous section,
the ``potential'' terms constructed by the spatial curvatures
grow exponentially in the Mixmaster case. 
These ``potential'' terms give rise to a infinite number of reflection and  
the spacetime will approach an oscillating state,
i.e. Mixmaster singularities described 
as an infinite sequence of Kasner epochs.  
On the other hand, 
such a ``potential'' term does not exist in the AVTD case.  
Therefore, reflection by the potential does not occur and the spacetime 
will approach a single Kasner epoch.  

Now, let us define AVTD.  
Suppose that a four-dimensional spacetime $(M,g)$ with the signature 
$(-+++)$
satisfies Einstein's equations, $G_{\mu\nu}=T_{\mu\nu}$, where 
$G_{\mu\nu}\equiv ^{\;(4)}\!\!\! R_{\mu\nu}-\frac{1}{2}^{\;(4)}\! Rg_{\mu\nu}$ 
is the Einstein tensor and 
$T_{\mu\nu}$ is the energy-momentum tensor.  
For any $3+1$ decomposition, we can obtain the constraint equations, 
\begin{equation}
^{(3)}\! R-K^{ab}K_{ab}+(tr K)^{2}=2T^{0}_{0},
\end{equation}
\begin{equation}
^{(3)}\!\nabla_{a}K^{a}_{b}-^{(3)}\!\nabla_{b}(tr K)=-T^{0}_{b},
\end{equation}
and the evolution equations, 
\begin{equation}
\partial_{t}h_{ab}=-2NK_{ab}+{\cal L}_{{\bf N}}h_{ab},
\end{equation}
\begin{equation}
\partial_{t}K^{a}_{b}=N\left[ ^{(3)}\! R^{a}_{b}+(tr K)K^{a}_{b}\right]
-^{(3)}\! \nabla_{b}  ^{(3)}\! \nabla^{a}N+
{\cal L}_{{\bf N}}K^{a}_{b}+2N\left[T^{a}_{b}
+\frac{1}{2}h^{a}_{b}(T^{0}_{0}-T^{c}_{c})\right],
\end{equation}
where $h_{ab}$, $K_{ab}$ and $(tr K)$ are the first and second fundamental 
form and 
mean curvature of the $3$-dimensional spacelike hypersurface, respectively.  
Also, $^{(3)}\! \nabla$, $^{(3)}\! R_{ab}$ and $^{(3)}\!R$ 
are the spatial covariant derivative, 
spatial Ricci and scalar curvature, respectively.  
$N$ and ${\bf N}$ are the lapse function and shift vector.  
${\cal L}_{{\bf N}}$ is the Lie derivative along the vector field ${\bf N}$.  
From the full Einstein equations, we
define the velocity term dominated (VTD) equations as follows:  
\begin{equation}
-K^{ab}K_{ab}+(tr K)^{2}=2\tilde{T}^{0}_{0},
\end{equation}
\begin{equation}
^{(3)}\! \nabla_{a}K^{a}_{b}-^{(3)}\! \nabla_{b}(tr K)=-\tilde{T}^{0}_{b},
\end{equation}
\begin{equation}
\partial_{t}h_{ab}=-2NK_{ab},
\end{equation}
\begin{equation}
\partial_{t}K^{a}_{b}=N\left[(tr K)K^{a}_{b}\right]
+2N\left[\tilde{T}^{a}_{b}
+\frac{1}{2}h^{a}_{b}(\tilde{T}^{0}_{0}-\tilde{T}^{c}_{c})\right],
\end{equation}
where $\tilde{\bullet}$ is modification of $\bullet$.  
(Generally, the spatial derivatives are removed.)  

In this setting we can give a definition of an AVTD spacetime and its 
singularity.  
\begin{definition}[Isenberg and Moncrief~\cite{IM}]
\label{def-avtd}
A spacetime $(M,g)$ is AVTD if 
\begin{enumerate}
\item it is a solution to Einstein's equations;
\item there exists another $(M,\tilde{g})$ and there exists a foliation 
$i_{t}:\Sigma^{3}\rightarrow M$ such that 
\begin{enumerate}
\item $(M,\tilde{g})$ obeys the VTD equations relative to $i_{t}$;
\item metric $g$ asymptotically approaches $\tilde{g}$ as $t\rightarrow 0$ 
in the sense that, for an appropriate norm $\|\bullet\|$ on the space of 
$3+1$ quantities $\bullet$, for any $\epsilon >0$ 
there exists $T>0$ such that
$\|\bullet - \tilde{\bullet}\|<\epsilon$ for all $t<T$. 
\end{enumerate}
\end{enumerate}
Furthermore, a spacetime $(M,g)$ has the AVTD singularity if 
\begin{enumerate}
\item it contains a spacelike singularity at $t\rightarrow 0$;
\item it is AVTD.  
\end{enumerate}
\end{definition}
To sum up, a singularity is called AVTD if 
all spatial curvature and spatial derivative terms in Einstein's equations 
can be neglected near the singularity 
$t\rightarrow 0$.  
Then, 
near the AVTD singularity, AVTD solutions can be interpreted as a different 
spatially homogeneous cosmology at each point in space.  

In order to see an example of the AVTD singularity, let us consider 
vacuum Bianchi I (Kasner) spacetimes.  
Since these spacetimes are spatially homogeneous, 
there is no spatial derivative terms.  Furthermore, 
they are spatially flat, i.e. $^{(3)}\! R=0$.  
Therefore, the singularity is necessarily AVTD.  
On the other hand, 
Bianchi IX spacetimes are not AVTD but Mixmaster~\cite{BKL,CM} since the 
spatial curvature does not vanish and the ``potential'' term constructed by 
the 
spatial curvature cannot be neglected as approaching to the singularity 
(See Sec.~\ref{why-emda}).  

\section{Influence of matter fields}
\label{why-emda}
BKL assumed the vacuum spacetimes originally in the investigation
of the singularity.  
The reason for putting such assumption is illustrated as follows.  
Consider class A Bianchi 
spacetimes with the metric 
$$
ds^{2}=-dt^{2}+e^{2\Omega}(e^{2\beta})_{ij}\sigma^{i}\sigma^{j},
$$
where $e^{\Omega}$ is the averaged scale factor, 
$\beta=$
diag$(-2\beta_{+}$, $\beta_{+}+\sqrt{3}\beta_{-}$, $\beta_{+}-\sqrt{3}\beta_{-})$ 
is a traceless matrix that determines the anisotropy and  
$d\sigma^{i}={\epsilon^{i}}_{ik}\sigma^{j}\wedge\sigma^{k}$.  
For simplicity, take $\sigma^{i}=dx^{i}$, i.e. Bianchi I spacetimes.  
Then, $e^{3\Omega}=t$ and $\Omega^{2}=\beta_{+}^{2}+\beta_{-}^{2}$.  
The spacetimes have singularities at $\Omega\rightarrow -\infty$.  
The square of shear $\sigma$ is estimated as
$\sigma^{2}\sim 
t^{-2}\sim e^{-6\Omega}.
$  
Further, consider a perfect fluid whose equation of state is 
$p=(\gamma -1)\rho$, where 
$\rho$ and $p$ are the energy density and the pressure, respectively.  
The contribution from the shear is more dominant than
that of the energy density  $\rho\sim e^{-3\Omega\gamma}$
near singularities ($\Omega\rightarrow -\infty$) if $1 \leq \gamma<2$ for 
natural matters such as a dust ($\gamma=1$) and a 
radiation  ($\gamma=4/3$).  
Thus, the effect of the matter field is negligible for the
behavior of the spacetime near singularities, and 
we have only to consider the vacuum case.   
Indeed, the known Bianchi perfect fluid solutions have similar behavior as 
the vacuum solutions near the singularities~\cite{WE}.  

We must, however, look more carefully into  
the influence of the energy density in the case
$\gamma =2$, where time dependence of the energy density
and shear are contribute equally. The matter with $\gamma=2$ is a stiff 
matter~\cite{JB}, and is not realistic matter in the relativistic sense. 
However, it is known that a scalar field $\phi$ is algebraically equivalent 
to a 
stiff matter if $\nabla\phi$ is timelike~\cite{WE}, where $\nabla$ is 
covariant derivative with respect to $g$.  
Therefore, the existence of scalar fields will give strong influence
on the nature of singularities.

Actually, Belinskii and Khalatnikov found that a massless scalar field can 
suppress Mixmaster 
oscillations~\cite{BK}.  
The result obtained by Belinskii and Khalatnikov 
is that all of Kasner exponents can be positive in 
Bianchi I spacetimes if the scalar field exists.  
Therefore, complicated permutation in the direction of anisotropic contraction 
toward singularities does not appear.  
Numerically, 
Berger confirmed Belinskii and Khalatnikov's result in Bianchi IX spacetimes 
and 
furthermore, she showed that Mixmaster 
oscillations are suppressed in 
non-polarized $U(1)$-symmetric and magnetic Gowdy spacetimes which are 
Mixmaster if there exist no scalar fields~\cite{BB00}.  

On the contrary, there is a possibility that 
the existence of exponential potentials of scalar fields drastically changes 
the behavior of singularities, i.e. the Mixmaster dynamics may come back.  
Recall the mechanism of the Mixmaster dynamics.   
The Hamiltonian ${\cal H}_{I\! X}$ 
for Bianchi IX spacetimes \cite{CM,BB00} is 
$$
2{\cal H}_{I\! X}=-p_{\Omega}^{2}+p_{+}^{2}+p_{-}^{2}+V(\Omega,\beta_{\pm}),
$$
where $p_{\Omega}$ and $p_{\pm}$ are conjugate 
momenta of $\Omega$ and $\beta_{\pm}$, respectively.  
The ``potential'' term $V(\Omega,\beta_{\pm})$ is given by 
$$
V(\Omega,\beta_{\pm})=e^{4\Omega} \left[e^{-8\beta_{+}}+e^{4
\beta_{+}+4\sqrt{3}\beta_{-}}+e^{4\beta_{+}-4\sqrt{3}\beta_{-}}
-2\left(e^{4\beta_{+}}+e^{-2\beta_{+}-2\sqrt{3}\beta_{-}}
+e^{-2\beta_{+}+2\sqrt{3}\beta_{-}}\right) \right].
$$
In the Bianchi I spacetimes there is no such ``potential'' term.  
It is known that 
Bianchi IX spacetimes have a singularity as $\Omega\rightarrow -\infty$.  
If the spacetimes are AVTD, the ``potential'' term must vanish 
as $\Omega\rightarrow -\infty$.  
When $V(\Omega,\beta_{\pm})=0$, 
we have $\beta_{\pm}=\beta_{\pm}^{0}+v_{\pm}|\Omega |$, 
$p_{\Omega}=$const. and $p_{\pm}=$const. 
by varying 
the Hamiltonian ${\cal H}_{I\! X}$, 
where $v_{\pm}\equiv  p_{\pm}/|p_{\Omega}|$ and $\beta_{\pm}^{0}$ are 
constant.  
The Hamiltonian constraint 
${\cal H}_{I\! X}=0$ is $v^{2}_{+}+v^{2}_{-}=1$.  
Then, 
$$
V\approx e^{-4|\Omega|(1+2\cos\theta)}
+e^{-4|\Omega|(1-\cos\theta-\sqrt{3}\sin\theta)}
+e^{-4|\Omega|(1-\cos\theta+\sqrt{3}\sin\theta)},
$$
where we parameterize $v_{+}=\cos\theta$ and $ v_{-}=\sin\theta$.  
For generic $\theta$ (except for $\theta=0,2\pi/3,4\pi/3$), 
the term $V$ exponentially grows as $\Omega\rightarrow -\infty$.  
This is a contradiction to vanish the ``potential'' term 
as $\Omega\rightarrow -\infty$.  
Thus, Bianchi IX spacetimes are not AVTD since the ``potential'' term becomes 
dominant near the singularity.  
As a result, reflections by the potential cause the Mixmaster dynamics.

As we have seen in the above discussion, 
the essential point is the existence of exponentially 
growing ``potential'' terms.   
Then, Belinskii and Khalatnikov have claimed that a scalar field $\phi$ exponentially 
coupled with the Maxwell field
restores Mixmaster oscillations in homogeneous spacetimes~\cite{BK}.  
Berger also suggested that if there exists a ``potential'' 
\begin{equation}
\label{potential}
V(\phi)
={\cal A}^{2}e^{\alpha\phi}+{\cal B}^{2}e^{-\beta\phi},
\hspace{.5cm}\alpha,\beta>0,
\end{equation}
where ${\cal A}$ and ${\cal B}$ are functions, then 
Mixmaster oscillations can be retained in both homogeneous and 
inhomogeneous spacetimes~\cite{BB00}.

We wish to explore how ``potential'' terms such as those in Eq.~(\ref{potential})
produced by matter fields might change the nature of the approach to
the singularity. As a first example we consider matter coupled to the
Gowdy spacetime. It has been shown mathematically and numerically
that vacuum Gowdy spacetimes are AVTD~\cite{BB98,GM,KR}. It is known
numerically that certain types of magnetic fields can ruin this
behavior~\cite{WIB}. While heuristically one would not expect this to
happen for the matter fields we study here, no rigorous demonstration
that Gowdy coupled to matter fields including
exponential potentials remains AVTD has been possible until now.

It should be noted that this form of potential arises
from the low energy effective superstring theory.
It contains the dilaton which is 
non-minimally coupled with the Maxwell and axion field.
From the unified theoretical point of view,
it is believed that the distinction between gravity and matter 
fields cannot be maintained near spacetime singularities and the superstring 
theory is the most promising candidate of such unified theories~\cite{STW}.  
Also, it is pointed out that the EMD system with a 
positive cosmological constant has new mechanism forming singularities
~\cite{MTN}.  
Hence, it is important to investigate the behavior around the
singularity in the EMDA system.

\section{Einstein-Maxwell-Dilaton-Axion system in Gowdy spacetimes 
on $T^{3}\times R$}
\label{emda-gowdy}

In this section, we introduce a standard model with additional 
assumptions. First,
we adopt Gowdy spacetimes.
Gowdy  proposed spatially compact and inhomogeneous 
spacetimes with the following metric~\cite{G}; 
\begin{equation}
\label{gowdy-metric}
ds^{2}=e^{\lambda(t,\theta)/2}t^{-1/2}(-dt^{2}+d\theta ^{2})+
R(t,\theta)\left[e^{-Z(t,\theta)}\left(dy+X(t,\theta)dz\right)^{2}
+e^{Z(t,\theta)}dz^{2}\right].
\end{equation}
Gowdy spacetimes have two twist free spacelike Killing vectors 
$\partial/\partial y$ and $\partial/\partial z$.  
The spatial topology of the spacetimes is classified into three types, 
$T^{3}$, $S^{2}\times S^{1}$ and $S^{3}$.  
We will consider only the simplest case, $T^{3}$, since the isometry group 
action has no degenerate orbits and therefore Einstein's equations have no 
corresponding singularities~\cite{VM81}.  
Gowdy spacetimes are called {\it polarized} if $X\equiv 0$, and 
{\it non-polarized} if this condition is not meet.

Properties of the metric (\ref{gowdy-metric}) depend on whether 
$\nabla R$ is timelike, spacelike or null.  
When the metric Eq.~(\ref{gowdy-metric}) describes a cosmological 
model, i.e. $\nabla R$ is globally 
timelike and the spatial topology is $T^{3}$ (periodic in $\theta$), 
one can take the function $R(t,\theta)=t$ without loss of generality 
by Gowdy's corner theorem if 
the spacetime is vacuum~\cite{G}.  
This fact be seen from Einstein's equations, 
\begin{equation}
\label{G-G=0}
G_{tt}-G_{\theta\theta}=\ddot{R}-R''=0,
\end{equation}
where dot and prime denote $t$ and $\theta$ derivatives, respectively.  
Also in the EDA system, Eq.~(\ref{G-G=0}) holds~\cite{JBK}.  
In the EM system, Eq.~(\ref{G-G=0}) is not satisfied
generically. However in the case that the field 
strength $F_{\mu\nu}=\partial_{\mu}A_{\nu}-\partial_{\nu}A_{\mu}$ 
has only the following components~\cite{PM};
$$
F_{ty}=\dot{\omega}, \;\;\;
F_{\theta y}=\omega ', \;\;\;
F_{tz}=\dot{\chi}, \;\;\;
F_{\theta z}=\chi ',
$$
Eq.~(\ref{G-G=0}) is satisfied.
This situation does not change even if the dilaton field is exponentially 
coupled with the Maxwell field.  
Thus, in the EMDA system, we can put $R(t,\theta)=t$.
Hereafter, we will choose this gauge.  
In this case, 
Gowdy spacetimes have spacelike singularities at $t=0$~\cite{G}.  

The action of the EMDA theory is 
\begin{eqnarray}
\label{action}
S&=&\int d^{4}x\sqrt{-g}\left[-\;^{(4)}\!R+e^{-2a\phi}F^{2}
+2\left(\nabla\phi\right)^{2}
+\frac{1}{3}e^{-4a\phi}H^{2}\right] \nonumber \\
&=& -\frac{1}{2}t\left[\ddot{\lambda}-\lambda ''+\dot{Z}^{2}-Z'^{2}
+e^{-2Z}\left(\dot{X}^{2}-X'^{2}\right)\right] \nonumber \\
&& -2e^{-2a\phi}\left[\left(e^{-Z}X^{2}+e^{Z}\right)
\left(\dot{\omega}^{2}-\omega '^{2}\right)
+e^{-Z}\left(\dot{\chi}^{2}-\chi '^{2}\right)
-2Xe^{-Z}\left(\dot{\omega}\dot{\chi}-\omega '\chi '\right)\right] \nonumber \\
&& -2t\left(\dot{\phi}^{2}-\phi '^{2}\right)
-\frac{1}{2}e^{4a\phi}t\left(\dot{\kappa}^{2}-\kappa'^{2}\right)
,
\end{eqnarray}
where $\phi =\phi (t,\theta)$ is the dilaton field, 
$H=dB\equiv -\frac{1}{2}e^{4a\phi}*d\kappa(t,\theta)$ is the three-index 
antisymmetric tensor field dual to the axion field $\kappa$, and 
$a$ is a coupling constant~\cite{STW}.  
The functions are periodic in $\theta$ ($0\leq \theta \leq 2\pi$) because 
the spatial topology is $T^{3}$.  
[However, the spatial compactness is not relevant to our analysis.]  
In the case of the effective superstring theory, $a=1$.
If we neglect the axion field, $a=\sqrt{3}$ gives
Kaluza-Klein theory. If $a=0$, the dilaton and axion
fields are decoupled with the gravity.

Note that the metric function $\lambda$ is decoupled with other functions, 
and that $\lambda$ appears only in the Hamiltonian and momentum constraints.  
This is the primary advantage of Gowdy spacetimes.  
Now, let us focus on evolution equations.  
We can calculate the metric function $\lambda$ by evaluating the integral of 
$\lambda '$ from $0$ to $2\pi$  after obtaining other functions.

\section{Review of the Fuchsian algorithm}
\label{Fuchsian algorithm}

The above system is very complicated and gives rise to
highly nonlinear partial differential equations 
which are nontrivially coupled with the matter fields.
We would like to construct AVTD singular solutions of such complicated 
equations which are parametrized by as 
many arbitrary functions as possible.  
In order to do this, we adopt the 
Fuchsian algorithm developed by Kichenassamy and Rendall~\cite{KR}.  
One of the advantages of this algorithm is applicable to 
general 
(nonlinear and/or singular) partial differential equations (PDEs).  
Another advantage of the Fuchsian algorithm is applicable to 
arbitrary spatial dimension.  
[Spatial dimension in Gowdy spacetimes is one since these spacetimes have two 
spacelike Killing vector fields.]
This fact suggests that the Fuchsian algorithm is applicable to 
spacetimes with fewer or no symmetry~\cite{R,LA}.  

In this section, we briefly review the Fuchsian algorithm~\cite{KR,IK,K}.  
Let us consider a PDE system  
\begin{equation}
\label{f=0}
F\left[u(t,x^{\alpha})\right]=0.
\end{equation}
Generically, $u$ can have any number of components.  Here, we will assume 
that the PDE is singular with respect to the argument $t$.  
The Fuchsian algorithm consists of three steps as follows:  


{\it Step 1}
Identify the leading (singular) 
terms $u_{0}(t,x)$ which are parts of the desired 
expansion for $u$.  This means that the most singular terms cancel each other 
when $u_{0}(t,x)$ is substituted in Eq.~(\ref{f=0}).  

{\it Step 2}
Introduce a renormalized unknown function $v(t,x)$, which is given by 
\begin{equation}
\label{u=a+tv}
u=u_{0}+t^{m}v.
\end{equation}
If $u_{0}\sim t^{k}$, we should set $m=k+\varepsilon$, 
where $\varepsilon >0$.  
Thus, $v$ is a regular part of the desired 
expansion for $u$.  

{\it Step 3}
Obtain a {\it Fuchsian system} for $v$ by substituting Eq.~(\ref{u=a+tv}) 
in Eq.~(\ref{f=0}).  That is, 
\begin{equation}
\label{fuchsian_system}
(D+A)v=t^{\epsilon}f(t,x,v,\partial_{x}v), 
\end{equation}
where $D\equiv t\partial _{t}$ and $A$ is a matrix which is independent of 
$t$ and $\epsilon >0$.  
$f$ can be assumed to be analytic in all of arguments except $t$ 
and continuous in $t$.  
Note that Eq.~(\ref{fuchsian_system}) is a ({\it singular}) PDE system for the 
{\it regular} function $v$.  

Roughly speaking, the Fuchsian algorithm is a transformation from 
finding singular 
solutions of the original equations (\ref{f=0}) into finding 
regular solutions of the Fuchsian equations (\ref{fuchsian_system}) which may 
be singular even if the original equations 
are regular.  
Once we have the Fuchsian system, we can show the existence of a unique 
solution for prescribed singular part $u_{0}$ by the following theorem.  
\begin{theorem}[Theorem 3 of Ref.~\cite{KR}]
\label{fuchsian_theorem}
Let us consider a system Eq.~(\ref{fuchsian_system}), 
where $A$ is an analytic matrix near $x=0$, such that $\|\sigma^{A}\|\leq C$ 
for $0<\sigma <1$ {\rm (boundedness condition)} 
and $f$ is analytic in space $x$ and continuous in time $t$.  
Then the system (\ref{fuchsian_system}) has exactly one solution which is 
defined near $x=0$ and $t=0$, and which is analytic in space and continuous 
in time, and tend to zero as $t\rightarrow 0$.  
\hfill$\Box$
\end{theorem}

Applying the Fuchsian algorithm to our problem 
of Gowdy spacetimes in the EMDA system, the procedure is summarized
as follows:  
\begin{enumerate}
\item Solve VTD equations as ordinary differential equations with respect to 
$t$.  
Solutions obtained such a way are the leading (singular) terms of the desired 
formal solutions.  
\item Substitute the formal solutions 
into full field equations.  
Here, make integral constants arbitrary functions of spatial arguments $x$.  
\item Obtain a Fuchsian system for unknown functions and evaluate every 
eigenvalues of $A$.  
\item Apply Theorem~\ref{fuchsian_theorem} to the system.  
\end{enumerate}
If every eigenvalues of $A$ are non-negative, the boundedness condition in 
Theorem~\ref{fuchsian_theorem} holds.  
Then, renormalized unknown functions must vanish as $t\rightarrow 0$ by 
Theorem~\ref{fuchsian_theorem}.  
Therefore, the only singular terms which are solutions to VTD 
equations remain and they are solutions to full field equations at $t=0$.  
Thus, we have AVTD singular solutions.  
Essentially, the above argument is a singular version of the 
Cauchy-Kowalevskaya theorem~\cite{LA,KR}.

\section{Asymptotic behavior in Gowdy spacetimes}
\label{results}

We are now ready to examine the singularity in the EMDA system.
The EMDA system with Gowdy symmetry, however, 
has six functions ($Z$, $X$, $\omega$, $\chi$, $\phi$, $\kappa$), and  
it is very complicated to treat all of them at the same time.  
Hence, we will reduce the system by selecting four functions
from them for simplicity.  
As a selection rule, it is important to leave exponential coupling between 
the dilaton and other fields since, as mentioned in Sec.\ref{why-emda}, 
an essential part of our analysis is whether or not 
``potential'' 
terms in Einstein's equations exponentially grow as $t\rightarrow 0$
~\cite{BB98,BB00}.  
Therefore, we will consider three cases, which are the non-polarized EMD, 
the non-polarized EDA and the polarized EMDA.

\subsection{non-polarized EMD  ($\chi =0$, $\kappa =0$)}
\label{sec-emd}
First, let us consider the case that the dilaton field is exponentially 
coupled with the Maxwell field.  
Varying the 
action (\ref{action}) with respect to ($Z$, $X$, $\omega$, $\chi$), 
we obtain evolution equations, 
\begin{eqnarray}
\label{emd-ee-z}
D^{2}Z-t^{2}Z'' + t^{2}e^{-2Z}(\dot{X}^{2}-X'^{2})
 + 2te^{-2a\phi}(e^{-Z}X^{2}-e^{Z})(\dot{\omega}^{2}-\omega'^{2})=0,
\\
\label{emd-ee-x}
D^{2}X-t^{2}X'' -2t^{2}(\dot{X}\dot{Z}-X'Z')
 - 4te^{-2a\phi +Z}X(\dot{\omega}^{2}-\omega'^{2})=0,
\\
\label{emd-ee-omega}
D^{2}\omega-t^{2}\omega''
-t^{2}\left[ 2a\dot{\phi}+\frac{1}{t}
-\frac{(2\dot{X}-\dot{Z}X)Xe^{-Z}
+\dot{Z}e^{Z}}{e^{-Z}X^{2}+e^{Z}}\right]
\dot{\omega} \hspace{1.1cm}\nonumber \\
+t^{2}\left[ 2a\phi '-\frac{(2X'-Z'X)Xe^{-Z}+Z'e^{Z}}{e^{-Z}X^{2}
+e^{Z}}\right]\omega'=0,
\\
\label{emd-ee-phi}
D^{2}\phi-t^{2}\phi ''
+ate^{-2a\phi}\left(e^{-Z}X^{2}+e^{Z}\right)(\dot{\omega}^{2}-\omega'^{2})=0.
\end{eqnarray}
Following the procedure explained in the previous section,
we have VTD equations by dropping spatial derivative terms 
from Eqs.(\ref{emd-ee-z})-(\ref{emd-ee-phi}).  
\begin{eqnarray}
\label{emd-vtd-z}
D^{2}Z+t^{2}e^{-2Z}\dot{X}^{2}
 +2te^{-2a\phi}(e^{-Z}X^{2}-e^{Z})\dot{\omega}^{2}=0, 
\\
\label{emd-vtd-x}
D^{2}X-2t^{2}\dot{X}\dot{Z}-4te^{-2a\phi +Z}X\dot{\omega}^{2}=0,
\\
\label{emd-vtd-omega}
D^{2}\omega
-t^{2}\left\{ 2a\dot{\phi}+\frac{1}{t}
-\frac{(2\dot{X}-\dot{Z}X)Xe^{-Z}
+\dot{Z}e^{Z}}{e^{-Z}X^{2}+e^{Z}}\right\}
\dot{\omega}
=0,
\\
\label{emd-vtd-phi}
D^{2}\phi+ate^{-2a\phi}(e^{-Z}X^{2}+e^{Z})\dot{\omega}^{2}=0.
\end{eqnarray}
According to Definition~\ref{def-avtd}, solutions to these equations are 
AVTD if they exist and are also solutions to full Einstein's 
equations as $t\rightarrow 0$.  

Since we expect Kasner-like solutions,  $Z\approx Z_{0}\ln t$, 
$X\approx X_{0}$ and $\phi\approx \phi_{0}\ln t$ are chosen 
as the leading terms for $Z$, $X$ and $\phi$, respectively.  
Substituting them into Eq.~(\ref{emd-vtd-omega}), we have a leading term of 
$\omega$.  
Thus, we can construct a family of formal solutions 
to full Einstein's equations as,  
\begin{eqnarray}
\label{emd-keisiki-z}
Z=Z_{0}(\theta)\ln t+Z_{1}(\theta)+t^{\epsilon}\alpha(t,\theta),
\\
\label{emd-keisiki-x}
X=X_{0}(\theta)+t^{2Z_{0}}\left(X_{1}(\theta)+\beta(t,\theta)\right),
\\
\label{emd-keisiki-omega}
\omega=\omega_{0}(\theta)+t^{Z_{0}+2a\phi _{0}+1}\left(\omega _{1}(\theta)
+\gamma(t,\theta)\right),
\\
\label{emd-keisiki-phi}
\phi=\phi _{0}(\theta)\ln t+\phi _{1} (\theta)+t^{\epsilon}\delta(t,\theta),
\end{eqnarray}
where $Z_{0}>0$, $Z_{0}+2a\phi _{0}+1>0$, $\epsilon>0$.  
The first and second inequalities are given by Eqs.~(\ref{emd-vtd-x}) 
and ~(\ref{emd-vtd-omega}).  
$\alpha$, $\beta$, $\gamma$ and $\delta$ are renormalized unknown functions.  
Note that when we put $\alpha =\beta =\gamma =\delta =0$,  
Eqs.~(\ref{emd-keisiki-z})-(\ref{emd-keisiki-phi}) are exact 
solutions to VTD equations as $t\rightarrow 0$.  
Substituting the formal solutions 
(\ref{emd-keisiki-z})-(\ref{emd-keisiki-phi}) 
into full Einstein's equations (\ref{emd-ee-z})-(\ref{emd-ee-phi}), 
we can obtain the 
following system,  
\begin{equation}
(D+A)\vec{u}=\vec{f}(t,\theta,\vec{u},\vec{u}'),
\label{fu}
\end{equation}
where 
\begin{eqnarray}
\vec{u} &=& 
(\alpha ,\; D\alpha ,\; t\alpha ',\; \beta ,\; D\beta ,\; t\beta ',
\; \gamma ,\; D\gamma ,\; 
t\gamma ',\; \delta ,\; D\delta ,\; t\delta')^{T}.
\label{uu}
\end{eqnarray}
The matrix $A$ has the $12\times 12$ components, 
$$
A=
\left(\begin{array}{cccc}
A_{Z} & 0 & 0 & 0 \\
0 & A_{X} & 0 & 0 \\
0 & 0 & A_{\omega} & 0 \\
0 & 0 & 0 & A_{\phi} \\
\end{array}\right),
$$
where,
$$
A_{Z}=
\left(\begin{array}{ccc}
0 & -1 & 0 \\
\epsilon^{2} & 2\epsilon & 0 \\
0 & 0  & 0 \\
\end{array}\right)
, \;\;\;\;
A_{X}=
\left(\begin{array}{ccc}
0 & -1 & 0 \\
0 & 2Z_{0} & 0 \\
0 & 0  & 0 \\
\end{array}\right),
$$
$$
A_{\omega}=
\left(\begin{array}{ccc}
0 & -1 & 0 \\
0 & Z_{0}+2a\phi _{0}+1 & 0 \\
0 & 0  & 0 \\
\end{array}\right)
,\;\;\;\;
A_{\phi}=
\left(\begin{array}{ccc}
0 & -1 & 0 \\
\epsilon^{2} & 2\epsilon & 0 \\
0 & 0  & 0 \\
\end{array}\right).
$$
We can easily verify that 
eigenvalues of $A$ are $0$, $2Z_{0}$, $Z_{0}+2a\phi _{0}+1$ and $\epsilon$, 
which are non-negative.  
Then, we conclude that the boundedness condition of 
Theorem~\ref{fuchsian_theorem} holds.  
Next, we examine the leading parts in components of the vector $\vec{f}$.  
$$
\vec{f}=
\left(\begin{array}{c}
0 \\
X_{0}'^{2}
e^{-2(Z_{1}+t^{\epsilon}\alpha)}t^{2-2Z_{0}-\epsilon} + 
2X_{0}^{2}\omega_{0}'^{2}
e^{-(2a(\phi_{1}+t^{\epsilon}\delta)+Z_{1}+t^{\epsilon}\alpha)}
t^{1-Z_{0}-2a\phi _{0}-\epsilon}+  \cdots \\
t(\alpha+D\alpha)' \\
0 \\
-2X_{0}'(Z_{0}'\ln t+Z_{1}'+t^{\epsilon}\alpha)t^{2-2Z_{0}}
-4X_{0}\omega_{0}'^{2}
e^{-2a(\phi_{1}+t^{\epsilon}\delta)+Z_{1}+t^{\epsilon}\alpha}
t^{1-Z_{0}-2a\phi _{0}} + \cdots \\
t(\beta+D\beta)' \\
0 \\
X_{0}^{-2}e^{Z_{1}+t^{\epsilon}\alpha}
\{\omega_{0}''-[2a(\phi_{0}'\ln t+\phi_{1}'+t^{\epsilon}\delta')+1]\omega_{0}'
\}t^{1-Z_{0}-2a\phi _{0}} + \cdots \\
t(\gamma+D\gamma)' \\
0 \\
aX_{0}^{2}\omega_{0}'^{2}
e^{-(2a(\phi_{1}+t^{\epsilon}\delta)+Z_{1}+t^{\epsilon}\alpha)}
t^{1-Z_{0}-2a\phi _{0}-\epsilon} + \cdots \\
t(\delta+D\delta)' \\
\end{array}\right).
$$
The dots represent the terms which are lower order with respect
to $t$ and are not important for our discussion since 
the power of these terms is necessarily positive if conditions $Z_{0}>0$ and 
$Z_{0}+2a\phi _{0}+1>0$ hold.  
Positivity of the power of $t$ in components of the vector $\vec{f}$ gives us 
sufficient conditions to obtain the 
AVTD solutions.  
Finally, we apply Theorem~\ref{fuchsian_theorem} to this system and then, 
we obtain the following theorem. 
\begin{theorem}
\label{th-nemd}
Let $Z_{i}(\theta)$, $X_{i}(\theta)$, $\omega_{i}(\theta)$, 
$\phi _{i}(\theta)$, where $i=0$, $1$, be real analytic functions and 
$0<Z_{0}<1$ and $-1<Z_{0}+2a\phi _{0}<1$ for $0\leq \theta \leq 2\pi$. 
Then, for field equations (\ref{emd-ee-z})-(\ref{emd-ee-phi}) in the 
EMD system, 
there exists a unique solution with the form 
(\ref{emd-keisiki-z})-(\ref{emd-keisiki-phi}), where 
$\alpha$, $\beta$, $\gamma$ and $\delta$ tend to zero as 
$t\rightarrow 0$.  
\hfill$\Box$
\end{theorem}
Theorem~\ref{th-nemd} means that $T^{3}$ Gowdy spacetimes in the 
EMD system have 
AVTD singularities in ``general'' in the sense that the singular solutions 
have maximal number (i.e. eight) of arbitrary functions ($Z_{0}$, $X_{0}$, 
$\omega_{0}$, 
$\phi _{0}$, $Z_{1}$, $X_{1}$, $\omega_{1}$, 
$\phi _{1}$).  

In components of the vector $\vec{f}$, terms $t^{1-Z_{0}-2a\phi _{0}}$ and 
$t^{2-2Z_{0}}$ give us upper bound for $Z_{0}$ and $Z_{0}+2a\phi _{0}$ in 
Theorem~\ref{th-nemd}.  However, since these terms are always 
multiplied by $X_{0}'$ and 
$\omega_{0}'$, the upper bound can be removed if $X_{0}'=0$ and 
$\omega_{0}'=0$.  
Hence, we obtain the following corollary: 

\begin{corollary}
\label{th-nemd'}
Let $Z_{i}(\theta)$, $X_{i}(\theta)$, $\omega_{i}(\theta)$, 
$\phi _{i}(\theta)$, where $i=0$, $1$, be real analytic functions, and 
assume $0<Z_{0}$, $-1<Z_{0}+2a\phi _{0}$ 
and $X_{0}'=\omega_{0}'=0$ for $0\leq \theta \leq 2\pi$. 
Then, for field equations (\ref{emd-ee-z})-(\ref{emd-ee-phi}) in the 
EMD system, 
there exists a unique solution with the form 
(\ref{emd-keisiki-z})-(\ref{emd-keisiki-phi}), where 
$\alpha$, $\beta$, $\gamma$ and $\delta$ tend to zero as 
$t\rightarrow 0$.  
\hfill$\Box$
\end{corollary}

Note that this is a ``non-generic'' case in the sense that two of 
arbitrary eight functions are constant in $\theta$.  

\subsection{non-polarized EDA ($\omega =0$, $\chi =0$)}
Now, we will consider the case that the axion field is exponentially 
coupled with the dilaton field.  In the similar way to Sec.\ref{sec-emd}, 
we have evolution equations by varying the action~(\ref{action}) 
with respect to $(Z, X, \phi, \kappa)$.  
\begin{eqnarray}
\label{eda-ee-z}
D^{2}Z-t^{2}Z''+t^{2}e^{-2Z}(\dot{X}^{2}-X'^{2})=0,
\\
\label{eda-ee-x}
D^{2}X-t^{2}X''-2t^{2}(\dot{X}\dot{Z}-X'Z')=0,
\\
\label{eda-ee-phi}
D^{2}\phi-t^{2}\phi ''-\frac{a}{2}t^{2}e^{4a\phi}(\dot{\kappa}^{2}-
\kappa '^{2})=0,
\\
\label{eda-ee-kappa}
D^{2}\kappa-t^{2}\kappa ''+4at^{2}(\dot{\phi}\dot{\kappa}-\phi '\kappa ')=0.
\end{eqnarray}
Note that a pair of PDEs for gravity (\ref{eda-ee-z}) and (\ref{eda-ee-x}) 
is decoupled with those
for scalar fields (\ref{eda-ee-phi}) and (\ref{eda-ee-kappa}).  
Furthermore, Eqs.~(\ref{eda-ee-z}) and (\ref{eda-ee-x}) have equivalent 
form to 
Eqs.~(\ref{eda-ee-phi}) and (\ref{eda-ee-kappa}) as simultaneous PDEs.  
These facts are known as ``mirror images'' in string cosmologies
and used to construct new nontrivial solutions~\cite{L}.  

Dropping the spatial derivative terms from 
Eqs.~(\ref{eda-ee-z})-(\ref{eda-ee-kappa}), 
VTD equations are obtained as follows:  
\begin{eqnarray}
\label{eda-vtd-z}
D^{2}Z+t^{2}e^{-2Z}\dot{X}^{2}=0,
\\
\label{eda-vtd-x}
D^{2}X-2t^{2}\dot{X}\dot{Z}=0,
\\
\label{eda-vtd-phi}
D^{2}\phi-\frac{a}{2}t^{2}e^{4a\phi}\dot{\kappa}^{2}=0,
\\
\label{eda-vtd-kappa}
D^{2}\kappa+4at^{2}\dot{\phi}\dot{\kappa}=0.
\end{eqnarray}
We can easily solve these equations as $t\rightarrow 0$ and find 
formal solutions to full Einstein's equations.  
\begin{eqnarray}
\label{eda-keisiki-z}
Z=Z_{0}(\theta)\ln t+Z_{1}(\theta)+t^{\epsilon}\alpha(t,\theta),
\\
\label{eda-keisiki-x}
X=X_{0}(\theta)+t^{2Z_{0}}(X_{1}(\theta)+\beta(t,\theta)),
\\
\label{eda-keisiki-phi}
\phi=\phi _{0}(\theta)\ln t+\phi _{1}(\theta)+t^{\epsilon}\gamma(t,\theta),
\\
\label{eda-keisiki-kappa}
\kappa=\kappa_{0}(\theta)+t^{-4a\phi _{0}}(\kappa _{1}(\theta)+\delta(t,\theta)),
\end{eqnarray}
where $Z_{0}>0$, $-2a\phi_{0} >0$ and $\epsilon>0$.  These restrictions for $Z_{0}$ 
and $\phi_{0}$ given by Eqs.~(\ref{eda-vtd-x}) and (\ref{eda-vtd-kappa}), 
respectively.  

The system for regular functions $\alpha$, $\beta$, $\gamma$ and $\delta$ 
in the EDA system is given by Eqs.~(\ref{fu})
and (\ref{uu}) by replacing the $12 \times 12$ matrix $A$ by
$$
A=
\left(\begin{array}{cccc}
A_{Z} & 0 & 0 & 0 \\
0 & A_{X} & 0 & 0 \\
0 & 0 & A_{\phi} & 0 \\
0 & 0 & 0 & A_{\kappa} \\
\end{array}\right),
$$
where 
$$
A_{\kappa}=
\left(\begin{array}{ccc}
0 & -1 & 0 \\
16a^{2}\phi_{0}^{2} & -4a\phi_{0} & 0 \\
0 & 0  & 0 \\
\end{array}\right).
$$
Eigenvalues of $A$ are $0$, $2Z_{0},-4a\phi _{0}$ and $\epsilon$, and all of 
them are non-negative.  These facts imply that the boundedness condition in 
Theorem~\ref{fuchsian_theorem} holds.  
After long calculation, 
the leading parts in components of the vector $\vec{f}$ is obtained as 
follows.  
$$
\vec{f}=
\left(\begin{array}{c}
0 
\\
X_{0}'^{2}e^{-2(Z_{1}+t^{\epsilon}\alpha)}t^{2-2Z_{0}-\epsilon}+ \cdots 
\\
t(\alpha+D\alpha)' 
\\
0 
\\
\left[X_{0}''-2X_{0}' (Z_{0}'\ln t+Z_{1}'+t^{\epsilon}\alpha')\right]
t^{2-2Z_{0}}+ \cdots 
\\
t(\beta+D\beta)'
\\
0 
\\
-\frac{a}{2}\kappa_{0}'^{2}e^{4a(\phi_{1}+t^{\epsilon}\gamma)}
t^{2+4a\phi _{0}-\epsilon} +\cdots 
\\
t(\gamma+D\gamma)' 
\\
0 
\\
\left[\kappa_{0}''+4a\kappa_{0}'(\phi_{0}'\ln t+\phi_{1}'
+t^{\epsilon}\gamma')\right]
t^{2+4a\phi _{0}}+ \cdots 
\\
t(\delta+D\delta)' 
\\
\end{array}\right).
$$
Since the power of $t$ of 
other terms in $\vec{f}$ is positive under conditions 
$Z_{0}>0$ and $-2a\phi_{0}>0$, and then
they are not needed here.  
The AVTD behavior requires that 
the power of the leading parts of $t$ is positive.  
Thus, we have the following theorem which states that non-polarized 
Gowdy spacetimes in the EDA system are AVTD in general.  
\begin{theorem}
\label{th-neda}
Let $Z_{i}(\theta)$, $X_{i}(\theta)$, $\phi _{i}(\theta)$, 
$\kappa_{i}(\theta)$, where $i=0$, $1$, be real analytic and 
$0<Z_{0}<1$ and $0<-2a\phi _{0}<1$ for $0\leq \theta \leq 2\pi$.  
Then, for field equations (\ref{eda-ee-z})-(\ref{eda-ee-kappa}) in the 
EDA system, 
there exists a unique solution of the form (\ref{eda-keisiki-z})-
(\ref{eda-keisiki-kappa}), 
where $\alpha$, $\beta$, $\gamma$ and $\delta$ tend to zero as $t\rightarrow 0$.  
\hfill$\Box$
\end{theorem}

As seen from the vector $\vec{f}$, the 
upper bound for $Z_{0}$ and $-\phi_{0}$ in 
Theorem~\ref{th-neda} are derived from terms multiplied by $X_{0}'$ and 
$\kappa_{0}'$.  Thus, we obtain the following corollary 
in the ``non-generic'' case.  

\begin{corollary}
\label{th-neda'}
Let $Z_{i}(\theta)$, $X_{i}(\theta)$, $\phi _{i}(\theta)$, 
$\kappa_{i}(\theta)$, where $i=0$, $1$, 
be real analytic, and assume
$0<Z_{0}$, $0<-2a\phi _{0}$ and $X_{0}'=\kappa_{0}'=0$,  
for $0\leq \theta \leq 2\pi$.  
Then, for field equations (\ref{eda-ee-z})-(\ref{eda-ee-kappa}) in the 
EDA system, 
there exists a unique solution of the form 
(\ref{eda-keisiki-z})-(\ref{eda-keisiki-kappa}), 
where $\alpha$, $\beta$, $\gamma$ and $\delta$ tend to zero as $t\rightarrow 0$.  
\hfill$\Box$
\end{corollary}
\subsection{polarized EMDA ($X=0$, $\chi =0$)}
Finally, we will study the case of polarized Gowdy spacetimes in 
the EMDA system.  
As mentioned in Sec.~\ref{why-emda}, the existence of ``potential'' like 
Eq.~(\ref{potential}) might lead to Mixmaster oscillations.  
Such a potential appears in the EMDA system [See Eq.~\ref{emda-ee-phi}].  
On the other hand, it has been shown that 
vacuum polarized $T^{3}$ Gowdy spacetimes are AVTD~\cite{IM}.  
Which behavior is realized in the EMDA system, AVTD or Mixmaster?

Varying the action~(\ref{action}) with respect to four functions 
$(Z, \omega, \phi, \kappa)$, we have evolution equations as follows:  
\begin{eqnarray}
\label{emda-ee-z}
D^{2}Z-t^{2}Z''-2te^{-2a\phi+Z}(\dot{\omega}^{2}-\omega'^{2})=0,
\\
\label{emda-ee-omega}
D^{2}\omega+t^{2}\left(\dot{Z}-2a\dot{\phi}-\frac{1}{t}\right)\dot{\omega}
-t^{2}\omega''-t^{2}(-2a\phi '+Z')\omega'=0,
\\
\label{emda-ee-phi}
D^{2}\phi-t^{2}\phi ''+ate^{-2a\phi+Z}(\dot{\omega}^{2}-\omega'^{2})+
\frac{1}{2}at^{2}e^{4a\phi}(\dot{\kappa}^{2}-\kappa '^{2})=0,
\\
\label{emda-ee-kappa}
D^{2}\kappa -t^{2}\kappa ''+4at^{2}(\dot{\kappa}\dot{\phi}-\kappa '\phi ')=0.
\end{eqnarray}
Dropping the spatial derivative terms from 
Eqs.~(\ref{emda-ee-z})-(\ref{emda-ee-kappa}), following
VTD equations are obtained as follows:  
\begin{eqnarray}
\label{emda-vtd-z}
D^{2}Z-2te^{-2a\phi+Z}\dot{\omega}^{2}=0,
\\
\label{emda-vtd-omega}
D^{2}\omega+t^{2}\left(\dot{Z}-2a\dot{\phi}-\frac{1}{t}\right)\dot{\omega}=0,
\\
\label{emda-vtd-phi}
D^{2}\phi+ate^{-2a\phi+Z}\dot{\omega}^{2}+
\frac{1}{2}at^{2}e^{4a\phi}\dot{\kappa}^{2}=0,
\\
\label{emda-vtd-kappa}
D^{2}\kappa +4at^{2}\dot{\kappa}\dot{\phi}=0.
\end{eqnarray}
Similarly to the case of the non-polarized EDA system, 
we can solve these equations 
as $t\rightarrow 0$.  
Thus, a family of formal solutions of this system can be obtained as follows:  
\begin{eqnarray}
\label{pmda-keisiki-z}
Z=Z_{0}(\theta)\ln t+Z_{1}(\theta)+t^{\epsilon}\alpha(t,\theta),
\\
\label{pmda-keisiki-omega}
\omega=\omega_{0}(\theta)+t^{2a\phi_{0}-Z_{0}+1}(\omega_{1}(\theta)
+\beta(t,\theta)),
\\
\label{pmda-keisiki-phi}
\phi=\phi_{0}(\theta)\ln t+\phi_{1} (\theta)+t^{\epsilon}\gamma(t,\theta),
\\
\label{pmda-keisiki-kappa}
\kappa=\kappa_{0}(\theta)+t^{-4a\phi_{0}}(\kappa_{1}(\theta)+\delta(t,\theta)),
\end{eqnarray}
where $2a\phi_{0}-Z_{0}+1>0$, $-4a\phi_{0}>0$ and $\epsilon>0$.  
Eqs.~(\ref{emda-vtd-omega}) and (\ref{emda-vtd-kappa}) restrict the values of 
$Z_{0}$ and $\phi_{0}$.  
By substituting Eqs.~(\ref{pmda-keisiki-z})-(\ref{pmda-keisiki-kappa}) into 
Eqs.~(\ref{emda-ee-z})-(\ref{emda-ee-kappa}), 
the system in the polarized EMDA system is given
by Eqs.~(\ref{fu}) and (\ref{uu}) with $A$ given by 
$$
A=
\left(\begin{array}{cccc}
A_{Z} & 0 & 0 & 0 \\
0 & A_{\omega} & 0 & 0 \\
0 & 0 & A_{\phi} & 0 \\
0 & 0 & 0 & A_{\kappa} \\
\end{array}\right),
$$
where 
$$
A_{\omega}=
\left(\begin{array}{ccc}
0 & -1 & 0 \\
0 & 2a\phi_{0}-Z_{0}+1 & 0 \\
0 & 0  & 0 \\
\end{array}\right), \;\;\;
A_{\kappa}=
\left(\begin{array}{ccc}
0 & -1 & 0 \\
0 & -4a\phi_{0} & 0 \\
0 & 0  & 0 \\
\end{array}\right).
$$
We can easily evaluate the
eigenvalues of the matrix $A$ as 
$0$, $2a\phi_{0}-Z_{0}+1$, $-4a\phi_{0}$ and $\epsilon$.  
Since they are all non-negative, the boundedness condition in 
Theorem~\ref{fuchsian_theorem} hold.  
The leading parts in components of the vector $\vec{f}$ are as follows:  
$$
\vec{f}=
\left(\begin{array}{c}
0 \\
-2\omega_{0}'^{2}
e^{ 
-2a(\phi_{1}+t^{\epsilon}\gamma)+Z_{1}+t^{\epsilon}\alpha}
t^{1-2a\phi_{0}+Z_{0}-\epsilon} +
\cdots \\
t(\alpha+D\alpha)' \\
0 \\
\{\omega_{0}''
+\omega_{0}'[-2a(\phi_{0}'\ln t+\phi_{1}'
+t^{\epsilon}\gamma ')+Z_{0}'\ln t+Z_{1}'
+t^{\epsilon}\alpha ']\}t^{1-2a\phi_{0}+Z_{0}}
+ \cdots 
\\
t(\beta+D\beta)' \\
0 \\
a\omega_{0}'^{2}
e^{-2a(\phi_{1}+t^{\epsilon}\gamma)+Z_{1}+t^{\epsilon}\alpha}
t^{1-2a\phi_{0}+Z_{0}-\epsilon}
+\frac{a}{2}\kappa_{0}'^{2}
e^{4a(\phi_{1}+t^{\epsilon})}
t^{2+4a\phi_{0}-\epsilon} + \cdots \\
t(\gamma+D\gamma)' \\
0 \\
\left[\kappa_{0}''+4a\kappa_{0}'(\phi_{0}'\ln t +\phi_{1}'
+t^{\epsilon}\gamma ')\right]
t^{2+4a\phi_{0}} + 
\cdots \\
t(\delta+D\delta)' \\
\end{array}\right).
$$
We do not need other terms since their power is positive under the
conditions $2a\phi_{0}-Z_{0}+1>0$ and $-4a\phi_{0}>0$.  
Thus, we have again the theorem which states that polarized 
Gowdy spacetimes in the EMDA system are AVTD in general 
even if there exists a ``potential'' as Eq.~(\ref{potential}) in the 
system.   
This result is a negative answer of the question in Sec.~\ref{why-emda}.  
\begin{theorem}
\label{th-pemda}
Let $Z_{i}(\theta)$, $\omega_{i}(\theta)$, $\phi _{i}(\theta)$, 
$\kappa_{i}(\theta)$, where $i=0$, $1$, be real analytic, and 
$-1<2a\phi_{0}-Z_{0}<1$ and $-1<2a\phi_{0}<0$ for $0\leq \theta \leq 2\pi$.  
Then, for field equations (\ref{emda-ee-z})-(\ref{emda-ee-kappa}) 
in the EMDA system, 
there exists a unique solution of the form 
(\ref{pmda-keisiki-z})-(\ref{pmda-keisiki-kappa}), 
where $\alpha$, $\beta$, $\gamma$ 
and $\delta$ tend to zero as 
$t\rightarrow 0$.  
\hfill$\Box$
\end{theorem}

The leading terms in components of the vector $\vec{f}$ always include factors 
$\omega_{0}'$ and/or $\kappa_{0}'$.  Then, we can be weaken the conditions for 
$Z_{0}$ and $\phi_{0}$ as similar to the cases of the EMD and the 
EDA 
systems if 
we put $\omega_{0}'=\kappa_{0}'=0$.  
\begin{corollary}
\label{th-pemda'}
Let $Z_{i}(\theta)$, $\omega_{i}(\theta)$, $\phi _{i}(\theta)$, 
$\kappa_{i}(\theta)$, where $i=0$, $1$, be real analytic, and assume 
$\omega_{0}'=\kappa_{0}'=0$, $-1<2a\phi_{0}-Z_{0}$ and $2a\phi_{0}<0$ 
for $0\leq \theta \leq 2\pi$.  
Then, for field equations (\ref{emda-ee-z})-(\ref{emda-ee-kappa}) 
in the EMDA system, 
there exists a unique solution of the form (\ref{pmda-keisiki-z})-
(\ref{pmda-keisiki-kappa}), where $\alpha$, $\beta$, $\gamma$ 
and $\delta$ tend to zero as 
$t\rightarrow 0$.  
\hfill$\Box$
\end{corollary}
\section{Summary}
\label{summary}
Theorem~\ref{th-nemd}, \ref{th-neda} and \ref{th-pemda} 
state that $T^{3}$ 
Gowdy spacetimes in the EMDA system have AVTD singularities in ``general'' 
in the sense that these 
AVTD singular solutions depend on the maximal number of singular data.  
In the ``non-generic'' cases where some functions in the singular data are 
constant in $\theta$, 
we have obtained corollary~\ref{th-nemd'}, \ref{th-neda'} 
and \ref{th-pemda'}.  
These results support BKL conjecture, that is, 
the dynamics of spatially different points effectively are decoupled from each 
other.  
However the spacetimes in our system do not show an oscillation 
of the Mixmaster spacetime near cosmological singularities
for some parameter regions.
In this sense, our system is a special case.

When we take the matter fields into account,
Theorem~\ref{th-pemda} and corollary~\ref{th-pemda'} give 
an answer, ``NO'', to the question in Sec.~\ref{why-emda}, i.e. 
the existence of a ``potential''  
Eq.~(\ref{potential}) does not necessarily restore the Mixmaster behavior.  
We must, however, impose another condition on the initial matter field 
in addition to the low-velocity condition to realize the
AVTD behavior by the effect of the  ``potential''.
Thus, we can say that 
the AVTD behavior is controlled by the ``potential''.

We can verify that 
the Kretschemann invariant $R^{\mu\nu\lambda\sigma}R_{\mu\nu\lambda\sigma}$ 
of all of our AVTD solutions blows up as
$t^{-(Z_{0}^{2}+2\phi_{0}^{2}+3)}$.  Thus, the AVTD singularity in the EMDA 
system is the curvature singularity.  
This supports the validity of the strong cosmic censorship for $T^{3}$ 
Gowdy spacetimes in the EMDA system since 
the spacetimes cannot be extended beyond the singularity similar to 
the vacuum case~\cite{GM,IM,VM81,CIM}.  

Let us compare our results with those obtained by 
Grubi\v{s}i\'{c}-Moncrief and 
Kichenassamy-Rendall in the (non-) polarized and 
vacuum case~\cite{GM,KR}.  
The low-velocity condition derived  by Grubi\v{s}i\'{c}-Moncrief and 
Kichenassamy-Rendall are equivalent to our condition 
$0<Z_{0}<1$.  If spacetimes are vacuum, then non-polarized $T^{3}$ Gowdy 
spacetimes are AVTD under this condition.  
Also, when $X\equiv 0$ (polarized), the spacetimes are AVTD without 
the low-velocity condition.  
In any case, our results clearly include Grubi\v{s}i\'{c}-Moncrief and 
Kichenassamy-Rendall's when the matter fields vanish.  

We should mention about some recent works related to our analysis.  
Weaver, Isenberg and Berger have shown numerically that  
a Gowdy spacetime in the EM system (a magnetic Gowdy spacetime) has the 
Mixmaster behavior~\cite{WIB}.  
The magnetic field they considered couples the metric function $\lambda$ 
through its amplitude.  
Our Maxwell fields, however, do not have such character (cf.~\cite{PM}), 
so Gowdy spacetimes in our EM system are able to show the AVTD behavior.  
It is expected that 
if Gowdy spacetimes in the EM system such as Weaver's model have Mixmaster behavior, 
they cannot show the AVTD even if we put the dilaton and the axion fields.

Andersson and Rendall have shown by using 
the Fuchsian algorithm that a generic cosmological 
spacetime in the Einstein-scalar field system has AVTD property~\cite{AR}.  
They have imposed the Gaussian coordinate conditions, $g_{00}=-1$ and 
$g_{0a}=0$ on the spacetime.  
As it is understood from its definition, AVTD is defined under a coordinate condition.  
Then, choice of gauge or coordinate conditions is important.  
Clearly, our gauge conditions (see Sec~\ref{emda-gowdy}) are different from 
Andersson-Rendall's.  
Furthermore, the equations of matter fields used in Ref.~\cite{AR} are {\it linear} PDEs.  
Contrary, our field equations in the EMDA system are {\it nonlinear}.  
These nonlinear PDEs cannot be came to linear ones.  
Thus, their and our results can be complements each of the other.  

Damour and Henneaux have shown under the Gaussian coordinate conditions 
that a generic cosmological singularity 
of low energy effective superstring theory is the Mixmaster~\cite{DH}.  
Their model is closely related to our model in the sense that it contains the matter 
fields with exponential potentials.  
However, there are some different points beside the gauge conditions : 
(i) the spacetime dimension (it is 11 or 10 in Ref.~\cite{DH}), 
(ii) an assumption to find formal solutions, and 
(iii) components of $p$-forms (i.e. Maxwell and axion) fields.  
It is known that the existence of the Mixmaster behavior strongly 
depends on the spacetime dimension.  
Also, the contributions of the $p$-form fields are 
neglected in the field equations when formal solutions, 
i.e. VTD solutions, are constructed.  
Then, these formal solutions are different from ours since 
the velocity terms of {\it any} fields are not neglected in our analysis.  
Furthermore, thanks to Gowdy symmetry in our analysis, 
components of $p$-form fields are restricted, that is, the Maxwell field strength can have 
only four components  
(see Sec.~\ref{emda-gowdy}), 
although there is no such restrictions in Ref.~\cite{DH}.  

To avoid complicated calculation we focused only on four functions among the 
six functions ($Z$, $X$, $\omega$, $\chi$, $\phi$ and $\kappa$)
in the present paper.  
For a next step, we would like to investigate the full system where
that all of six functions do not vanish.  
Another subject to consider is asymptotic behavior of 
$U(1)$-symmetric spacetimes~\cite{VM86} which 
have only one spacelike Killing vector.  
Although Einstein's equations for $U(1)$-symmetric spacetimes are more 
complicated than those of $T^3$ Gowdy spacetime,
the Fuchsian algorithm is applicable to such
system since this algorithm is independent on dimension of space.  
Recently, the Fuchsian algorithm is used for vacuum polarized $U(1)$-symmetric 
spacetimes~\cite{IK,LA}.  
Our interest is to know whether non-polarized $U(1)$-symmetric and 
generic cosmological spacetimes in 
the EMDA system are AVTD or not.  
Although the procedure of the Fuchsian algorithm is routine, 
finding out plausible formal solutions to VTD equations and obtaining the Fuchsian 
system are algebraically complicated, and not so easy.  
Therefore, these subjects are devoted to the future investigation.  

\begin{center}
{\bf Acknowledgment}
\end{center}
We are grateful to Akio Hosoya, Hideki Ishihara, Hideo Kodama, 
Kei-ichi Maeda and 
Shigeaki Yahikozawa for useful discussions.  
We also thank Michael Ashworth for his careful reading of our manuscript.  
This work is partially supported by Scientific Research 
Fund of the Ministry of Education, Science, Sports, and 
Culture, by the Grant-in-Aid for 
JSPS (No. 199704162 (T. T.) and No. 199906147 (K. M.)).

\end{document}